\documentclass[aps,prl,twocolumn,superscriptaddress]{revtex4}

\usepackage[latin9]{inputenc}
\usepackage{amsmath}
\usepackage{amssymb}
\usepackage{graphicx}
\usepackage{makecell}
\usepackage{array}
\usepackage{CJK}

\usepackage[unicode=true,pdfusetitle,
 bookmarks=false,
 breaklinks=false,pdfborder={0 0 1},backref=false,colorlinks=false]
 {hyperref}
\hypersetup{
 bookmarksnumbered=false,bookmarksopen=false}

\makeatletter
\@ifundefined{textcolor}{}
{%
 \definecolor{BLACK}{gray}{0}
 \definecolor{WHITE}{gray}{1}
 \definecolor{RED}{rgb}{1,0,0}
 \definecolor{GREEN}{rgb}{0,1,0}
 \definecolor{BLUE}{rgb}{0,0,1}
 \definecolor{CYAN}{cmyk}{1,0,0,0}
 \definecolor{MAGENTA}{cmyk}{0,1,0,0}
 \definecolor{YELLOW}{cmyk}{0,0,1,0}
}

\usepackage{color}

\newcommand{\ehbar}{\hbar_{\mathrm{eff}}}

\@ifundefined{textcolor}{}{%
 \definecolor{BLACK}{gray}{0}
 \definecolor{WHITE}{gray}{1}
 \definecolor{RED}{rgb}{1,0,0}
 \definecolor{GREEN}{rgb}{0,1,0}
 \definecolor{BLUE}{rgb}{0,0,1}
 \definecolor{CYAN}{cmyk}{1,0,0,0}
 \definecolor{MAGENTA}{cmyk}{0,1,0,0}
 \definecolor{YELLOW}{cmyk}{0,0,1,0}
}

\usepackage{txfonts}

\makeatother

\begin{document}

\title{Dynamical transition of quantum scrambling in a non-Hermitian Floquet synthetic system}

\author{Liang Huo}
\affiliation{School of Information Engineering, Jiangxi University of Science and Technology, Ganzhou 341000, China}

\author{Han Ke}
\affiliation{School of Science, Jiangxi University of Science and Technology, Ganzhou 341000, China}

\author{Wen-Lei Zhao}
\email[]{wlzhao@jxust.edu.cn}
\affiliation{School of Science, Jiangxi University of Science and Technology, Ganzhou 341000, China}

\begin{abstract}
We investigate the dynamics of quantum scrambling, characterized by the out-of-time ordered correlators (OTOCs), in a non-Hermitian quantum kicked rotor subjected to quasi-periodical modulation in kicking potential. Quasi-periodic modulation with incommensurate frequencies creates a high-dimensional synthetic space, where two different phases of quantum scrambling emerge: the freezing phase characterized by the rapid increase of OTOCs towards saturation, and the chaotic scrambling featured by the linear growth of OTOCs with time. We find the dynamical transition from the freezing phase to the chaotic scrambling phase, which is assisted by increasing the real part of the kicking potential along with a zero value of its imaginary part. The opposite transition occurs with the increase in the imaginary part of the kicking potential, demonstrating the suppression of quantum scrambling by non-Hermiticity. The underlying mechanism is uncovered by the extension of the Floquet theory. Possible applications in the field of quantum information are discussed.
\end{abstract}
\date{\today}

\maketitle

\section{Introduction}
Understanding the connection between scrambling and chaos is fundamentally important in different fields of physics, such as quantum information and quantum chaos~\cite{Dowling23prl,TXu20prl,ritzsch22pre,JWang21pre}.
It is found that the out-of-time ordered correlators (OTOCs) can be exploited as an effective tool to detect the quantum information scrambling, quantum chaos and quantum entanglement~\cite{LewisSwan19,XDHu23}. The underlying chaos with exponential instability generally gives rise to the fast scrambling of quantum information in many-body systems~\cite{Belyansky20prl,JKim21prb}, which is quantified by the exponential growth of OTOCs with the rate denoted by quantum Lyapunov exponent~\cite{Maldacena16,Rozenbaum17}. Interestingly, genuine quantum chaos induced by many-body interaction~\cite{ZQi23} and Floquet-engineered nonlinear interaction~\cite{WLZhao21prb} even induces the fastest scramblers, being characterized by the superexponential growth of OTOCs in the LMG model~\cite{ZQi23} and in the Gross-Pitaevskii map model~\cite{WLZhao21prb}. The rich physics represented by OTOCs' dynamics enlightens our understanding on the nonequilibrium dynamics of closed quantum many-body systems, such as quantum thermalization~\cite{Murthy19prl}, ergodicity~\cite{Rozenbaum19prb} and topological phase transitions~\cite{Bin23prb}, which is of interest for fundamental physics from both theoretical~\cite{Kukuljan17prb} and experimental viewpoints~\cite{Chaudhury09}.

Floquet engineering technique has now been widely employed to create versatile synthetic systems which have potential advantages to mimic the electronic transport in a nanotube threaded by an artificial gauge field~\cite{Hainaut18nc}, the Anderson metal-insulator transition in the high dimensional space~\cite{Lopez13}, as well as the topological fractional pumping of cold atoms~\cite{Taddia17prl}, just to name a few. It is more interesting that the complex potentials are experimentally achievable by using a decay channel of cold atoms to excited state in the standing laser field~\cite{Chudesnikov91,Keller97} and by taking account to the loss feature of optical medium~\cite{Berry04,Berry08,Yin13}. This non-Hermitian extension of conventional systems induces the flourishing advancement of novel platforms, such as non-Hermitian photonic crystal, non-Hermitian sensors~\cite{Budich20}, and non-Hermitian electrical circuits~\cite{Stegmaier21}, which have been widely used to explore the fundamental concepts of topology, degeneracies, and adiabaticity~\cite{WYWang22pra,XWang23njp} in non-Hermitian physics. For $\cal{PT}$-symmetric systems, the dynamics of OTOCs can diagnose Yang-Lee edge singularity~\cite{Zhai20prb}, exhibit the quantized response to external potential~\cite{Wlzhao22prr}, and display the scaling law at the transition to the spontaneous $\cal{PT}$-symmetry breaking~\cite{Wlzhao23pra}, which sheds light on the fruitful physics in Floquet synthetic systems.

In this context, we investigate the dynamics of OTOCs in a non-Hermitian kicked rotor (NQKR) model for which the complex kicking potential is modulated quasi-periodically in time. In this Floquet synthetic system, we observe a rapid increase of the OTOCs towards saturation during time evolution, when the real part of the kicking potential is small and its the imaginary part is zero. This demonstrates the freezing phase of quantum scrambling for long-term evolution behaviors. In this Hermitian case, a dynamical transition for the OTOCs from the freezing behavior to power law growth emerges as the real part of kicking potential increases, indicating the chaos assisted scrambling. Interestingly, the opposite transition from power-law behavior to freezing behavior occurs by increasing the imaginary part of the kicking potential. We further obtain the phase diagram of the quantum scrambling by utilizing a well-trained long short-term memory network (LSTM). By extending the Floquet theory, we predict that a quantum state will finally evolve to a quasi-eigenstate of the instantaneous Floquet operator with the large most value of the imaginary part of quasi-energy. This is verified by our numerical results of the fidelity between time evolved quantum state and quasi-eigenstates. The feature of exponential localization of quasi-eigenstates determines the freezing of quantum scrambling of the NQKR model.

The paper is organized as follows. In Sec.~\ref{MMResl}, we describe our model and show the dynamics of OTOCs. We discuss the mechanism of the freezing of quantum scrambling in Sec.~\ref{Mechasi}. Conclusion and discussions presented in Sec.~\ref{Concl}.

\section{Model and main results}\label{MMResl}
The dimensionless Hamiltonian of the QPKR model reads
\begin{equation}\label{Hamil}
{\rm H}=\frac{{p}^2}{2}+ V_K(\theta,t) \sum_n
\delta(t-t_n)\:,
\end{equation}
with the time-dependent kicking potential
\begin{equation}\label{NHKicking}
V_K(\theta,t)= (K+i\lambda)[1+\eta\cos(\omega_{1}t)\cos(\omega_{2}t)]\cos(\theta)\;,
\end{equation}
where $p=-i\ehbar\partial/\partial \theta$ is the angular momentum operator, $\theta$ is the angle coordinate, satisfying the commutation relation $[\theta,p]=i\ehbar$ with $\ehbar$ effective Planck constant. Here, the parameters $K$ and $\lambda$ control the strength of the real and imaginary parts of the kicking potential, respectively. The time $t_n$ is integer, i.e., $t_n = 1, 2\ldots$, indicating the kicking number. The $\omega_j$ ($j=1,2$) indicates the modulation frequency on the kicking strength, with the modulation strength denoted by $\eta$. When $\omega_1$ and $\omega_2$ are incommensurate with each other, and both of them are incommensurate with $2\pi$, this quasi-periodical extension of the QKR model mimics a three-dimensional QKR~\cite{Casati89,Shepelyansky83,Lopez13,Hainaut2018}. We consider the case with $\omega_{1}=2\pi/\kappa$ and $\omega_{2}=2\pi/\kappa^{2}$, where $\kappa=1.3247\ldots$ is the real root of the cubic equation~\cite{Shepelyansky14}.

The eigenequation of angular momentum operator is $p|n\rangle = p_n |n \rangle$ with eigenvalue $p_n = n\ehbar$ and eigenstate $\langle \theta |n\rangle=e^{in\theta}/\sqrt{2\pi}$. With this complete basis, an arbitrary state can be expanded as $|\psi \rangle=\sum_n \psi_n |n\rangle$.
The advancements of the delta-kicking systems is that the Floquet operator can be split into two components, namely $U(t)=U_fU_K(t)$, with the free evolution operator $U_f = \exp(-ip^2/2\ehbar)$ and the kicking term $U_K(t)=\exp[-iV_K(\theta,t)/2\ehbar]$. The time evolution of a quantum state from $t_n$ to $t_{n+1}$ is governed by $|\psi(t_{n+1})\rangle = U(t_{n})|\psi(t_{n})\rangle$.

The OTOCs are defined as $C(t)=-\langle[A(t),B]^2\rangle$, where both $A(t)=U^{\dagger}(t)A U(t)$ and $B$ are operators evaluated in Heisenberg picture, and $\langle \cdot\rangle=\langle \psi(t_0)|\cdot|\psi(t_0)\rangle$ denotes the average over an initial state. Since the thermal state of Floquet systems are not well defined, there is no need to perform the average over the ensemble of thermal state in evaluating the $C$~\cite{DAlessio14,Zhao21,Zhao22}. We consider the case that $A$ is the translation operator $A=e^{-i\epsilon p}$ and $B$ is the projection operator on the initial state $B=|\psi(t_0)\rangle\langle \psi(t_0)|$. Straightforward derivation yields the relation $C(t)=\mathcal{N}^2(t)-|\langle \psi(t)|e^{i\epsilon p}|\psi(t)\rangle|^2$ with the norm $\mathcal{N}(t)=\langle \psi(t)|\psi(t)\rangle$. The non-unitary evolution of non-Hermitian Hamiltonian leads to the exponential growth of the norm, i.e., $\mathcal{N}(t)=e^{\gamma t}$. In order to reduce the impact of the norm to OTOCs, we introduce the rescaled OTOCs as $C(t)=1-|\langle \psi(t)|e^{-i\epsilon p}|\psi(t)\rangle|^2/\mathcal{N}^2(t)$. In condition that $\epsilon \ll 1$, we can derive the approximate equivalence
\begin{equation}\label{OTCEng}
C(t)\approx \epsilon^2[\langle p^2(t)\rangle-\left(\langle p(t)\rangle\right)^2]\;,
\end{equation}
where we used the approximation $e^{-i\epsilon p}\approx 1-i\epsilon p$~\cite{Zhao23arx15,Zhao23arx62}. This relation clearly demonstrates the underlying connection between quantum scrambling and energy diffusion~\cite{Lewis19,Meier19}, which has practical applications in enhancing the precision of quantum measurement~\cite{ZYLi2022,Garttner18}. In numerical simulations, we choose the ground state of the angular momentum operator as initial state, i.e., $|\psi(t_0)\rangle =1/\sqrt{2\pi}$.

\begin{figure}[t]
\begin{center}
\includegraphics[width=8cm]{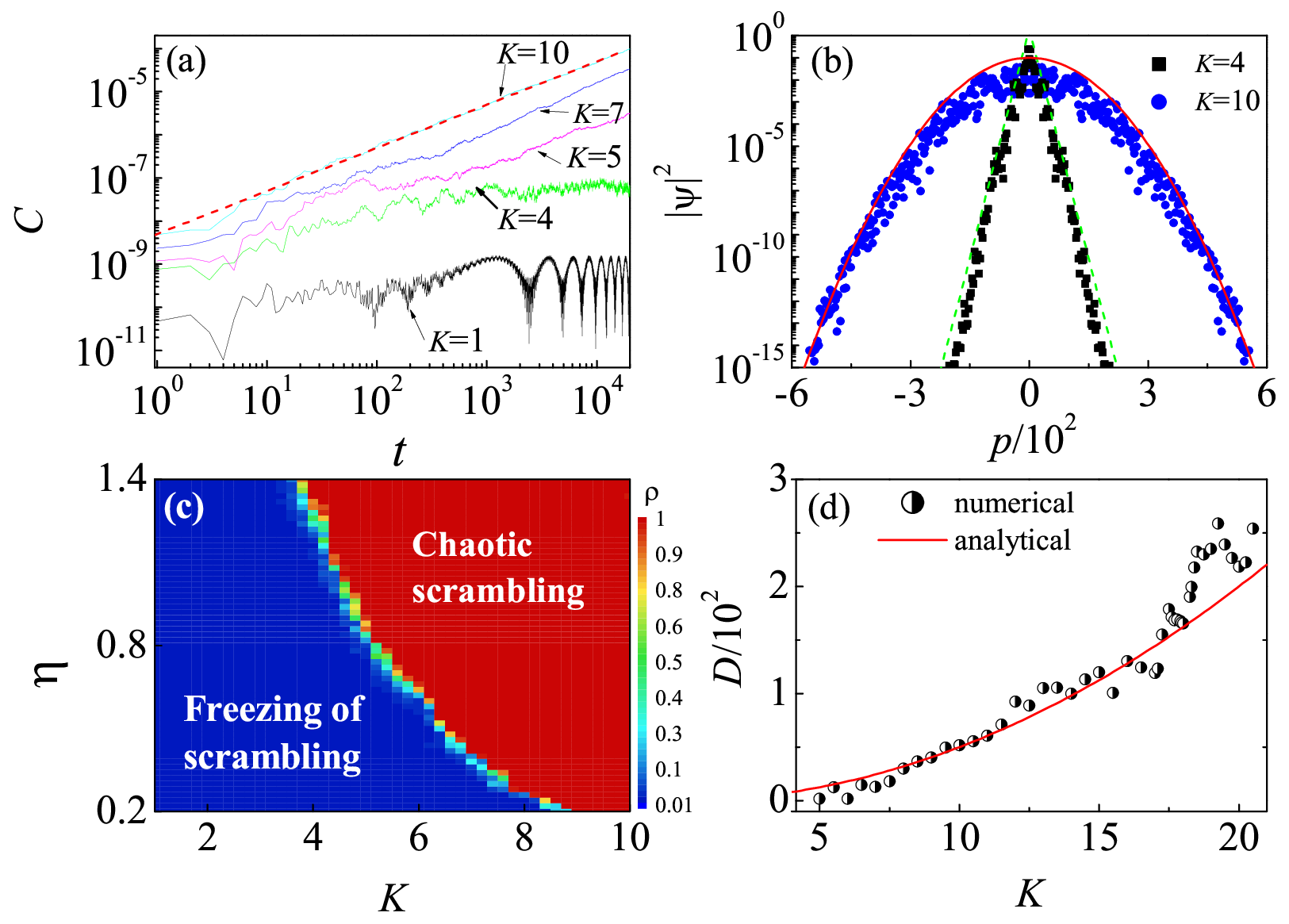}
\caption{(a) Time evolution of $C$ for $\lambda=0$ with $K=1$ (black line), 4 (green line), 5 (pink line), 7 (blue line), and 10 (cyan line). Red-dashed line indicates our theoretical prediction $C\approx (\epsilon K)^2t/2$. (b) Momentum distributions at the time $t= 1000$ for $K=4$ (squares) and 10 (circles). Red solid (green dashed) line indicates the Gaussian function (exponentially-localized function), i.e., $|\psi(p)|^2\propto \exp(-p^2/\sigma)$ ($|\psi(p)|^2\propto \exp(-|p|/\xi)$). (c) Phase diagram of the quantum scrambling in the parameter space $(\eta,K)$. One can see clearly a phase boundary. Here, the value of $\rho$ indicates the probability of a parameter value $(\eta,K)$ being within regime of the freezing of scrambling ($\rho=0$) or the chaotic scrambling ($\rho=1$). (d) The growth rate $D$ versus $K$. Solid line denotes our theoretical prediction ($D\approx K^2/2$). The parameters are $\eta=0.75$, $\epsilon = 10^{-5}$ and $\ehbar=2.89$.
}
\label{HermiOTC}
\end{center}
\end{figure}

We numerically investigate the time evolution of $C$ for a wide regime of $K$. Figure~\ref{HermiOTC}(a) shows that $C$ increases during short time duration, and eventually fluctuates around a saturation level for small $K$. The saturation level increases with the increase of $K$. This demonstrates the suppress of quantum scrambling by the dynamical localization of quantum diffusion. In fact, we have uncovered the power-law dependence of a different OTOCs on the kicking strength in the dynamical localization regime~\cite{Zhao23pra201}. Interestingly, the OTOCs linearly increases with time $C(t)=Dt$ when $K$ is sufficiently large (e.g., $K=7$ and 10), indicating the emergence of chaotic scrambling. Corresponding to the freezing of scrambling, the wavepackets are exponentially localized in momentum space, i.e., $|\psi(p)|^2\propto \exp(-|p|/\xi)$ with a constant $\xi$, which is a signature of the dynamical localization [see Fig.~\ref{HermiOTC}(b)]. For the linear-law of scrambling, the momentum distribution can be well described by the Gaussian function $|\psi(p)|^2\propto \exp(-p^2/\sigma)$, which characterizes the chaotic diffusion.

It is now widely accepted that the machine learning method can be used to determine the feature of the time sequence of observables by using finite time evolution of these observables~\cite{HZhao21}. Previously, we have use a well-trained LSTM network to successfully acquire the phase diagram of the spontaneous $\cal{PT}$-symmetry breaking in a QKR model~\cite{Zhao23pra201}.
Here, we use the LSTM network to explore the phase diagram of the quantum scrambling in parameter space $(\eta,K)$. For a specific value of $(\eta,K)$, it generate a probability $\rho$ for the finite time sequence of the $C(t)$, which is featured by the asymptotical saturation ($\rho=0$) or the linear growth ($\rho=1$). Figure~\ref{HermiOTC}(c) demonstrates a clear phase boundary $\eta_c$ separating the phase of the freezing of scrambling ($\rho=0$) and the chaotic scrambling phase ($\rho=1$). The critical value $\eta_c$ monotonically decreases with the increase of $K$. That is to say, the larger $K$ makes it easier for the transition from the freezing scrambling phase to chaotic scrambling occurs. In fact, the parameter $K$ controls the hopping distance in momentum-space lattice~\cite{Lopez13}. Therefore, strong kick strength is helpful for enhancing the quantum transport behavior. Since the $C$ is proportional to the variance of wavepackets in momentum space, the phase diagram of $C$ in the parameter in space $(\eta, K)$ is essentially same as that of the mean energy~\cite{Lopez13,Mano21}.

Since both $\omega_{1}$ and $\omega_{2}$ are irrational numbers in the time dependent kick strength, this kind of temporal modulation induces effectively random noises ($\eta \ll 1$) to kick strength in time domain. It is well-known that random noises in kick strength can destroy DL and lead to classically normal diffusion in the system, i.e., $\langle p^2\rangle \approx K^2t/2$~\cite{Lemarie10JMO}. Substituting this relation into Eq.~\eqref{OTCEng} yields $C(t)\approx (\epsilon K)^2t/2$, where we use the condition $\langle p\rangle =0$ as the quantum state is symmetric in momentum space. Therefore, the scrambling rate $D= dC/dt$ equals to $(\epsilon K)^2/2$, which is validated by our numerical results in Fig.~\ref{HermiOTC}(d).
\begin{figure}[t]
\begin{center}
\includegraphics[width=7cm]{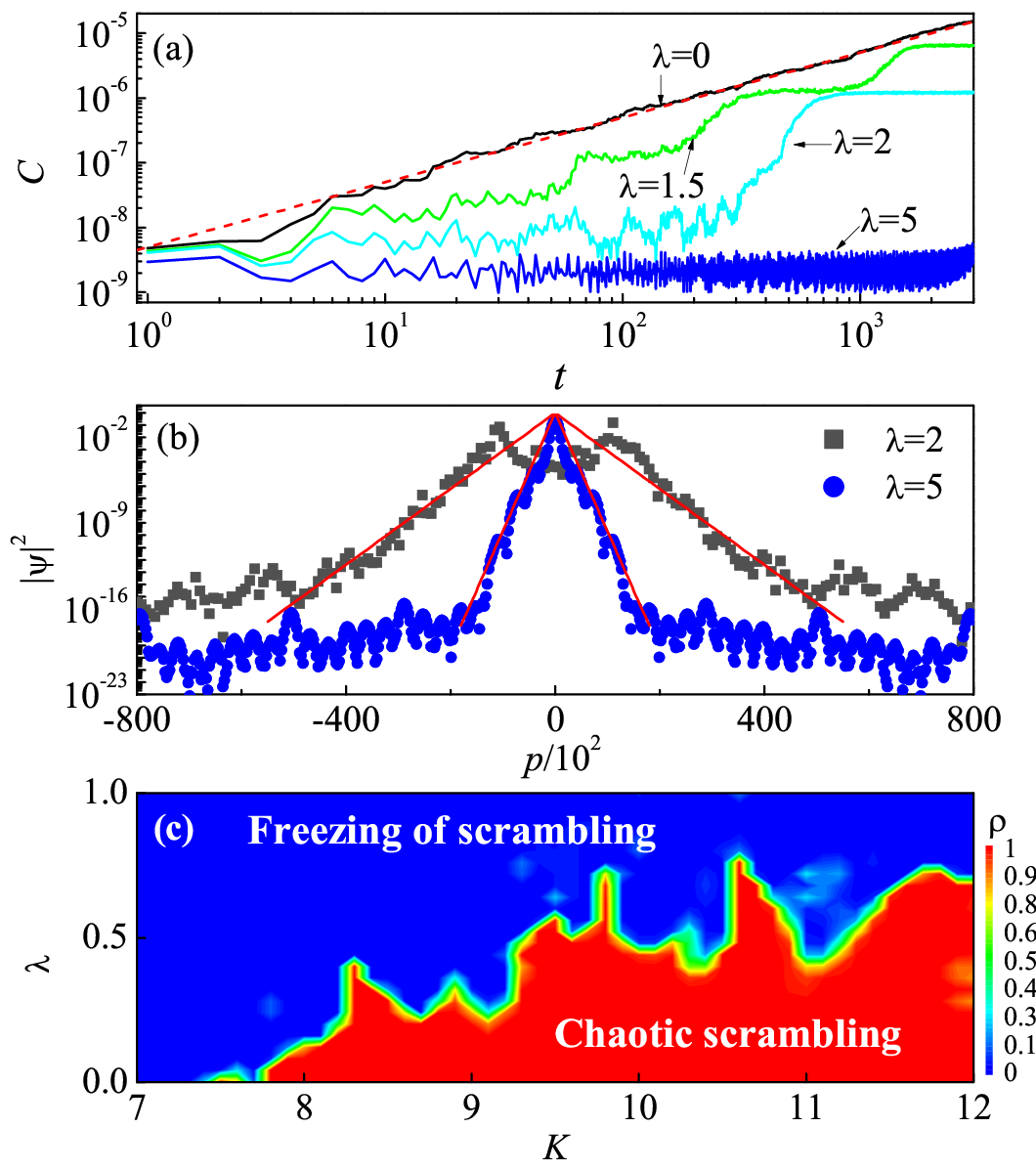}
\caption{(a) Dependence of $C$ on time with $K=10$ for $\lambda = 0$ (black line), 1.5 (green line), 2 (cyan line), and 5 (blue line).  Red-dashed line indicates the linear growth $C\approx (\epsilon K)^2t/2$. (b) Momentum distributions at the time $t=1000$ for $\lambda = 2$ (squares) and 5 (circles). Red solid lines denote the exponentially-localized function $|\psi(p)|^2\propto \exp(-|p|/\xi)$ with $\xi\approx 14$ and $4.5$ for $\lambda=2$ and 5, respectively. (c) Phase diagram of the quantum scrambling in the parameter space $(\lambda, K)$. Other parameters are the same as in Fig.~\ref{HermiOTC}(d).
}
\label{NonHQScramb}
\end{center}
\end{figure}

In order to reveal the effect of the non-Hermitian driven potential on the quantum scrambling dynamics, we numerically investigate the time dependence of $C$ for different $\lambda$. Figure~\ref{NonHQScramb}(a) shows that for $\lambda = 1.5$ the growth of the $C$ is reduced compared with the $C$ for $\lambda=0$, which demonstrates the suppress of the scrambling by non-Hermiticity. For sufficiently large $\lambda$ (e.g., $\lambda=2$ and $5$), the $C$ exhibits the asymptotical saturation after long time evolution, moreover the saturation level decreases with the increase of $\lambda$, indicating the total freezing of quantum scrambling for $\lambda \gg 1$. Correspondingly, the wavepackets are exponentially localized in momentum space,i.e., $|\psi(p)|^2\propto \exp(-|p|/\xi)$ with localization length $\xi$ decreasing with the increase of $\lambda$ [see Fig.~\ref{NonHQScramb}(a) for $\lambda = 5$], which reveals the mechanism of the suppress of quantum scrambling by dynamical localization. The decrease of $\xi$ with the increase of the $\lambda$ also reflects the enhancement of the dynamical localization by the non-Hermitian driven potential, whose mechanism has been uncovered by our previous investigations~\cite{Wlzhao23sym,Wywang22,KQHuang21pra}. We further use the LSTM network to investigate the dynamical transition of the $C$ in the parameter space $(\lambda,K)$. Figure~\ref{NonHQScramb}(c) shows that there is a clear boundary quantified by $\lambda_c$ between the freezing of scrambling ($\lambda > \lambda_c$) and the chaotic scrambling phases ($\lambda < \lambda_c$). This phase diagram broadens our understanding on the quantum scrambling in non-Hermitian Floquet systems.

\section{Mechanism of the freezing of scrambling in non-Hermitian regime}\label{Mechasi}

\begin{figure}[t]
\begin{center}
\includegraphics[width=7.5cm]{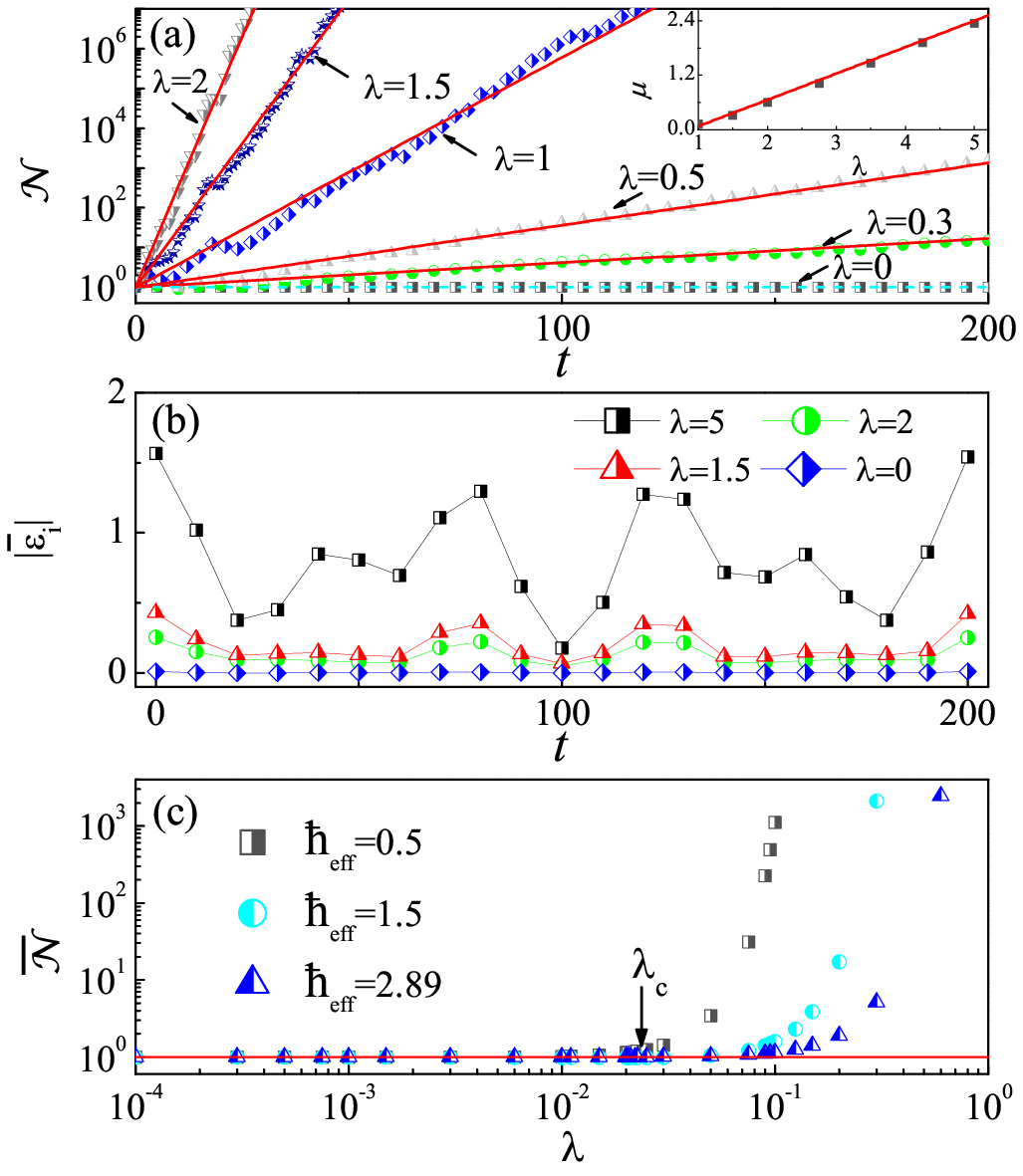}
\caption{(a) Norm $\mathcal{N}$ versus time for $\lambda = 0$ (squares), 0.3 (circles), 0.5 (up triangles), 1 (diamonds), 1.5 (pentagrams), and 2 (down triangles). Solid lines indicate the exponential increase, i.e., $\mathcal{N} = e^{\mu t}$. Dashed line denotes $\mathcal{N}=1$. Inset: The $\mu$ versus $\lambda$. Solid line indicates the linear growth $\mu \propto \lambda$. (b) Time dependence of the average value of the imaginary parts of the quasi-energies $|\bar{\varepsilon}_{i}|$ for $\lambda$ = 0 (diamonds), 1.5 (triangles), 2 (circles), 5 (squares). The parameters are $K=10$ and $\ehbar=2.89$. (c) Average value $\bar{\mathcal{N}}$ versus $\lambda$ for $K=10$ with $\ehbar = 0.5$ (squares), 1.5 (circles), and 2.89 (triangles). Arrow marks the critical value $\lambda_c$. Solid line denotes $\mathcal{N} = 1$. Other parameters are the same as in Fig.~\ref{HermiOTC}(d).}\label{Nonhbreaking}
\end{center}
\end{figure}

A character of the non-unitary evolution of non-Hermitian systems is the growth of the norm with time. We numerically investigate the time evolution of the norm $\mathcal{N}$ for different $\lambda$. Figure~\ref{Nonhbreaking}(a) shows that, for $\lambda$ smaller than a threshold value $\lambda_c$, the value of $\mathcal{N}$ equals almost to unity with time evolution corresponding to the real quasienergy. For $\lambda > \lambda_c$, the $\mathcal{N}$ increases exponentially with time, i.e., $\mathcal{N} = e^{\mu t}$, for which the growth rate $\mu$ linearly increases with the increase of $\lambda$ [see the inset in Fig.~\ref{Nonhbreaking}(a)]. For periodically-driven systems, the Floquet theory governs the eigenequation $U |\varphi_{\varepsilon}\rangle=e^{-i\varepsilon}|\varphi_{\varepsilon}\rangle$. In analogy with this equation, we define the eigenequation of the instantaneous Floquet operator as $U(t+1,t)|\varphi_{\varepsilon}(t)\rangle=e^{-i\varepsilon (t)}|\varphi_{\varepsilon}(t)\rangle$. It is reasonable to believe that the quasienergies are complex $\varepsilon=\varepsilon_r + i\varepsilon_i$ for non-unitary Floquet operator. To confirm this issue, we numerically investigate the time evolution of the average value of the imaginary part of the quasienergy $|\bar{\varepsilon}_i(t)|=\sum_{j=1}^{N}|\varepsilon_i^j(t)|/M$ where $M$ is the dimension of the Floquet operator matrix. Figure~\ref{Nonhbreaking}(b) shows that the value of $|\bar{\varepsilon}_i|$ is virtually zero for small $\lambda$, corresponding to $\mathcal{N}=1$ in Fig.~\ref{Nonhbreaking}(a). For sufficiently large $\lambda$ (e.g., $\lambda =2$), the $|\bar{\varepsilon}_i(t)|$ oscillates around a saturation level, which appearantly increases with $\lambda$. The appearance of complex quasienergies leads to the exponential growth of the norm. We further investigate the long-time average value of the norm $\bar{\mathcal{N}}$ for a wide range of $\lambda$.  Figure~\ref{Nonhbreaking}(c) shows that, for a specific $\ehbar$, the $\bar{\mathcal{N}}$ remains at unity for $\lambda$ smaller than a threshold value $\lambda_c$, and increases rapdily with the increase of $\lambda$ for $\lambda > \lambda_c$. The critical value $\lambda_c$ increases with the increase of $\ehbar$.

\begin{figure}[t]
\begin{center}
\includegraphics[width=8cm]{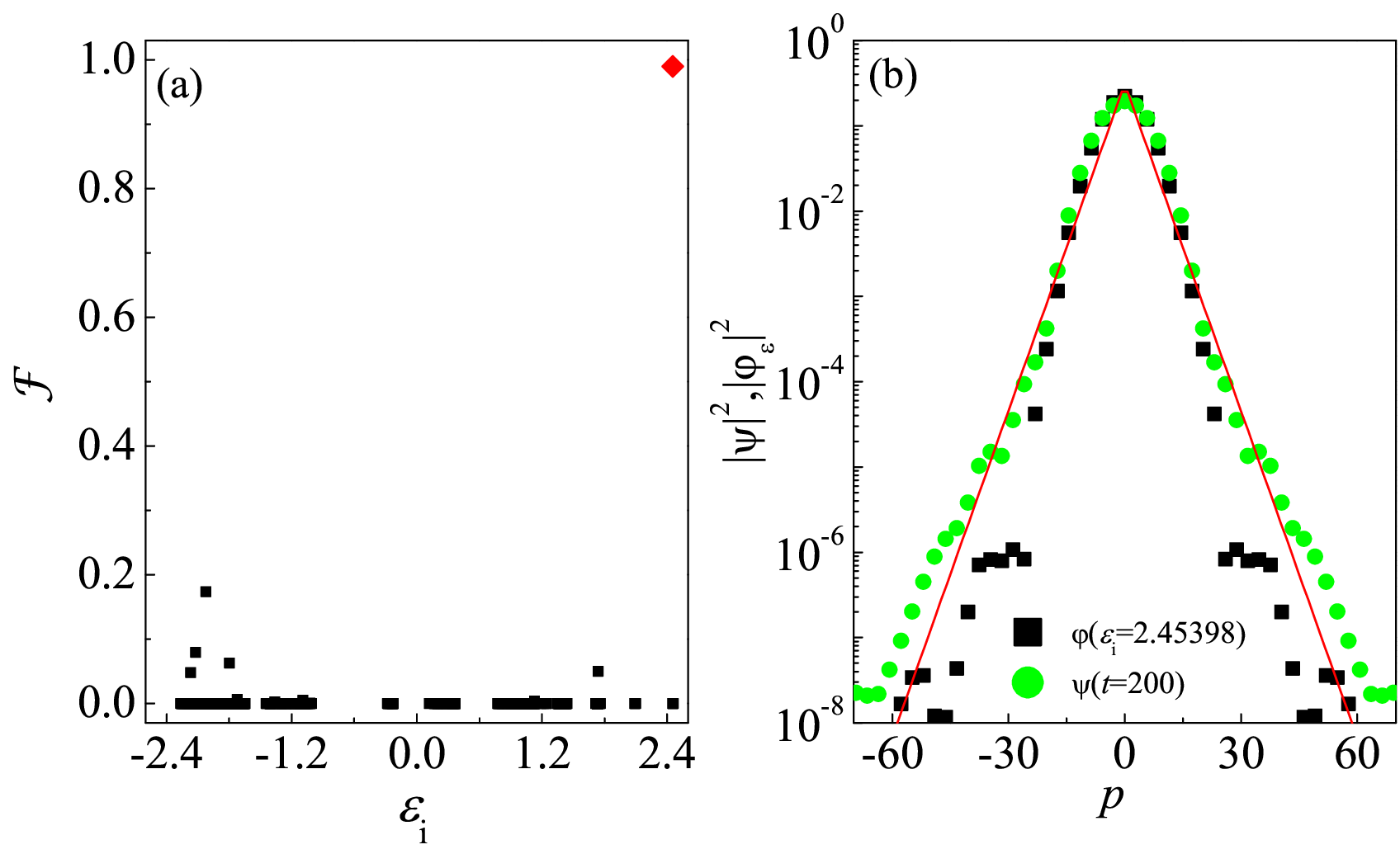}
\caption{(a) Fidelity $\mathcal{F}$ at the time $t = 200$ versus the imaginary part of the quasi-energy $\varepsilon_{i}$ for $K=10$ and $\lambda = 5$. Red diamond indicates $\mathcal{F}\approx 1$ for $\varepsilon_i= 2.45398$. (b) Comparison of the probability density distribution between the state $|\psi(t= 200)\rangle$ and the quasi-eigenstate $|\varphi_{\varepsilon}(t= 200)\rangle$ with $\varepsilon_i=2.45398$. Solid line indicates the exponentially-localized distribution of $|\psi(p)|^{2}\propto \exp(-|p|/\xi)$ with $\xi\approx 3.4$. Other parameters are the same as in Fig.~\ref{HermiOTC}(d).}\label{MechaFidel}
\end{center}
\end{figure}

We have found both theoretically and numerically that the quantum state of non-Hermitian Floquet systems will evolved to the quasi-eigenstate whose imaginary part of the quasi-energie is maximum. Thus, the feature of this state determines the dynamical behaviors of the system~\cite{Wlzhao23sym,Wywang22,KQHuang21pra,KQHuang21jpcm}.
As a futher step, we numerically investigate the fidelity between the time-evolved quantum state and the quasieigenstates, i.e., $\mathcal{F}(t) = |\langle\psi(t)|\varphi_{\varepsilon}(t)\rangle$ for $\lambda \gg \lambda_c$. Figure~\eqref{MechaFidel}(a) shows that for $\lambda = 5$, the $\mathcal{F}$ at the time $t = 200$ almost equals to unity at the maximum $\varepsilon_i = 2.45398$. In addition, we compare the probability density distribution between quantum state $|\psi(t = 200)\rangle$ and quasieigenstate $|\varphi_{\varepsilon}(t = 200)\rangle$ with $\varepsilon_i = 2.45398$ in Fig.~\eqref{MechaFidel}(a). One can find that the two states almost completely overlap with each other, both of which are exponentially localized in momentum space. It is apparent that the exponential localization of quasieigenstates suppresses the quantum scrambling.

\section{Conclusion and discussions}\label{Concl}

In the present work, we investigate the dynamics of quantum scrambling in high dimensional synthetic space created by quasi-periodical modulation on a NQKR model. In Hermitian case (i.e., $\lambda=0$), we find a dynamical transition of quantum scrambling from freezing phase to the chaotic scrambling of linear growth of OTOCs by increasing the real part $K$ of the kicking strength. By increasing the imaginary part of the kicking potential, the system demonstrates the dynamical transition from chaotic scrambling to freezing behavior, demonstrating the
suppress of the scrambling by non-Hermiticity. We obtain the phase diagram of the quantum scrambling with a kind of machine learning method, i.e., the LSTM. In order to reveal the dynamical transtion of the quantum scrambling, we extend the Floquet theory to our system, and find the emergence of complex quasi-energies of the instantaneous Floquet operator. The fidelity between time evolved quantum state with quasi-eigenstates reveals that the quantum state finally evolves to the quasi-eigenstate with large most value of the imaginary quasi-energy. Its localized property dominates the freezing of quantum scrambling. Our findings serve as new elements of the quantum scrambling in synthetic systems generated by Floquet engineering.

\section*{ACKNOWLEDGMENTS}

This work is supported by the National Natural Science Foundation of China (Grant No. 12065009 and 12365002), the Science and Technology Planning Project of Jiangxi province (Grant No. 20224ACB201006, 20224BAB201023, and 20212BAB204024).


\begin{thebibliography}{*}
\bibitem{Dowling23prl}
N. Dowling, P. Kos, and K. Modi, Scrambling Is Necessary but Not Sufficient for Chaos, Phys. Rev. Lett. {\bf 131}, 180403 (2023).

\bibitem{TXu20prl}
T. Xu, T. Scaffidi, and X. Cao, Does Scrambling Equal Chaos? Phys. Rev. Lett. {\bf 124}, 140602 (2020).

\bibitem{ritzsch22pre}
F. Fritzsch and T. Prosen, Boundary chaos, Phys. Rev. E {\bf 106}, 014210 (2022).

\bibitem{JWang21pre}
J. Wang, G. Benenti, G. Casati, and W. G. Wang, Quantum chaos and the correspondence principle, \pre {\bf 103}, L030201 (2021).


\bibitem{LewisSwan19}
R. J. Lewis-Swan, A. Safavi-Naini, J. J. Bollinger, and A. M. Rey, Unifying scrambling, thermalization and entanglement through measurement of fidelity out-of-time-order
correlators in the Dicke model, Nat. Commun. {\bf 10}, 1581 (2019).

\bibitem{XDHu23}
X. D. Hu, T. Luo, and D. B. Zhang, Quantum algorithm for evaluating operator size with Bell measurements, \pra {\bf107}, 022407 (2023).


\bibitem{Belyansky20prl}
R. Belyansky, P. Bienias, Y. A. Kharkov, A. V. Gorshkov, and B. Swingle, Minimal Model for Fast Scrambling,  Phys. Rev. Lett. {\bf125}, 130601 (2020).

\bibitem{JKim21prb}
J. Kim, E. Altman, and X. Cao, Dirac fast scramblers, Phys. Rev. B {\bf103}, L081113 (2021).

\bibitem{Maldacena16}
J. Maldacena, S. H. Shenker, and D. Stanford, A bound on chaos, J. High Energy Phys. {\bf106} (2016).

\bibitem{Rozenbaum17}
E. B. Rozenbaum, S. Ganeshan, and V. Galitski, Universal level statistics of the out-of-time-ordered operator, Phys. Rev. B {\bf 100}, 035112 (2019).

\bibitem{ZQi23}
Z. Qi, T. Scaffidi, and X. Cao, Surprises in the deep Hilbert space of all-to-all systems: From superexponential scrambling to slow entanglement growth, \prb {\bf 108}, 054301 (2023).

\bibitem{WLZhao21prb}
W. L. Zhao, Y. Hu, Z. Li, and Q. Wang, Super-exponential growth of Out-of-time-ordered correlators, Phys. Rev. B {\bf 103}, 184311 (2021).

\bibitem{Murthy19prl}
C. Murthy and M. Srednicki, Bounds on Chaos from the Eigenstate Thermalization Hypothesis, Phys. Rev. Lett. {\bf 123}, 230606 (2019).

\bibitem{Rozenbaum19prb}
E. B. Rozenbaum, S. Ganeshan, and V. Galitski, Universal level statistics of the out-of-time-ordered operator, Phys. Rev. B {\bf 100}, 035112 (2019).

\bibitem{Bin23prb}
Q. Bin, L. L. Wan, F. Nori, Y. Wu, and X. Y. L{\" u}, Out-of-time-order correlation as a witness for topological phase transitions, Phys. Rev. B {\bf 107}, L020202 (2023).

\bibitem{Kukuljan17prb}
I. Kukuljan, S. Grozdanov, and T. Prosen, Weak quantum chaos, Phys. Rev. B {\bf 96}, 060301(R) (2017).

\bibitem{Chaudhury09}
S. Chaudhury, A. Smith, B. E. Anderson, S. Ghose and P. S. Jessen, Quantum signatures of chaos in a kicked top. Nature {\bf461}, 768-771 (2009).

\bibitem{Hainaut18nc}
C. Hainaut, I. Manai, J.-F. Cl{\' e}ment, J. C. Garreau, P. Szriftgiser, G. Lemari{\' e}, N. Cherroret, D. Delande, and R. Chicireanu, Controlling Symmetry and Localization with an Artificial Gauge Field in a Disordered Quantum System, Nat. Commun. {\bf 9}, 1382 (2018).

\bibitem{Lopez13}
M. Lopez, J. F. Cl{\' e}ment, G. Lemari{\' e}, D. Delande, P. Szriftgiser and J. C. Garreau, Phase diagram of the anisotropic Anderson transition with the atomic kicked rotor: theory and experiment, New J. Phys. {\bf 15} 065013 (2013).

\bibitem{Taddia17prl}
L. Taddia, E. Cornfeld, D. Rossini, L. Mazza, E. Sela, and R. Fazio, Topological Fractional Pumping with Alkaline-Earth-Like Atoms in Synthetic Lattices, Phys. Rev. Lett. {\bf118}, 230402 (2017).


\bibitem{Chudesnikov91}
D. O. Chudesnikov and V. P. Yakovlev, Bragg scattering on complex potential and formation of super-narrow momentum distributions of atoms in light fields, Laser Phys. {\bf1}, 110 (1991).

\bibitem{Keller97}
C. Keller, M. K. Oberthaler, R. Abfalterer, S. Bernet, J. Schmiedmayer, and A. Zeilinger, Tailored Complex Potentials and Friedel's Law in Atom Optics, \prl {\bf 79}, 3327 (1997).


\bibitem{Berry04}
M. Berry, Physics of Nonhermitian Degeneracies, Czech. J. Phys. {\bf 54}, 1039 (2004).

\bibitem{Berry08}
M V Berry, Optical lattices with PT symmetry are not transparent, J. Phys. A: Math. Theor. {\bf41}, 244007 (2008).

\bibitem{Yin13}
X. Yin and X. Zhang, Unidirectional light propagation at exceptional points, Nat. Mater. {\bf 12}, 175 (2013).

\bibitem{Budich20}
J. C. Budich and E. J. Bergholtz, Non-Hermitian Topological Sensors, Phys. Rev. Lett. {\bf125}, 180403 (2020).

\bibitem{Stegmaier21}
A. Stegmaier, {\it et al.}, Topological defect engineering and $\cal{PT}$ symmetry in non-Hermitian electrical circuits, Phys. Rev. Lett. {\bf126}, 215302 (2021).

\bibitem{WYWang22pra}
W. Y. Wang, B. Sun, and J. Liu, Adiabaticity in nonreciprocal Landau-Zener tunneling, Phys. Rev. A {\bf106}, 063708 (2022).

\bibitem{XWang23njp}
X. Wang, H. D. Liu, and L. B. Fu, Nonlinear non-Hermitian Landau-Zener-St{\" u}ckelberg-Majorana
interferometry, New J. Phys. {\bf25} 043032 (2023).


\bibitem{Zhai20prb}
L. Zhai and S. Yin, Out-of-time-ordered correlator in non-Hermitian quantum systems, Phys. Rev. B {\bf102}, 054303 (2020).

\bibitem{Wlzhao22prr}
W. L. Zhao, Quantization of out-of-time-ordered correlators in non-Hermitian chaotic systems, Phys. Rev. Research {\bf4}, 023004 (2022).


\bibitem{Wlzhao23pra}
W. L. Zhao, R. R. Wang, H. Ke, and J. Liu, Scaling laws of the out-of-time-order correlators at the transition to the spontaneous $\cal{PT}$-symmetry breaking in a Floquet system, \pra {\bf 107}, 062201 (2023).





\bibitem{Casati89}
G. Casati, I. Guarneri, and D. L. Shepelyansky, Anderson Transition in a One-Dimensional System with Three Incommensurate Frequencies, Phys. Rev. Lett. {\bf 62}, 345 (1989).


\bibitem{Shepelyansky83}
D. L. Shepelyansky, Some statistical properties of simple classically stochastic quantum systems, Physica D: Nonlinear Phenomena {\bf 8} 208 (1983).


\bibitem{Hainaut2018}
C. Hainaut, P. Fang, A. Ran{\c {o}}on, J.-F. Cl{\'e}ment, P. Szriftgiser, J.-C. Garreau, C. S. Tian, and R. Chicireanu, Experimental Observation of a Time-Driven Phase Transition in Quantum Chaos, , Phys. Rev. Lett. {\bf 121}, 134101 (2018).


\bibitem{Shepelyansky14}
L. Ermann and D. L. Shepelyansky, Destruction of Anderson localization by nonlinearity in kicked rotator at different effective dimensions, J. Phys. A: Math. Theor. {\bf 47} 335101 (2014).

\bibitem{DAlessio14}
L. D'Alessio and M. Rigol, Long-time Behavior of Isolated Periodically Driven Interacting Lattice Systems, Phys. Rev. X {\bf 4}, 041048 (2014).

\bibitem{Zhao21}
W. L. Zhao, Y. Hu, Z. Li, and Q. Wang, Super-exponential growth of Out-of-time-ordered correlators, Phys. Rev. B {\bf 103}, 184311 (2021).

\bibitem{Zhao22}
W. L. Zhao, Quantization of Out-of-Time-Ordered Correlators in non-Hermitian Chaotic Systems, Phys. Rev. Research {\bf 4}, 023004 (2022).


\bibitem{Zhao23arx15}
W. L. Zhao and J. Liu, Superexponential behaviors of out-of-time ordered correlators and Loschmidt echo in a non-Hermitian interacting system, arXiv:2305.1215.


\bibitem{Zhao23arx62}
W. L. Zhao and J. Liu, Quantum criticality at the boundary of the non-Hermitian regime of a Floquet system, arXiv:2307.00462.

\bibitem{Lewis19}
R. J. Lewis-Swan, A. Safavi-Naini, J. J. Bollinger, and A. M. Rey, Unifying scrambling, thermalization and entanglement through measurement of fidelity out-of-time-order correlators in the Dicke model, Nat. Commun. {\bf 10}, 1581 (2019).

\bibitem{Meier19}
E. J. Meier, J. Ang'ong'a, F. Alex An, and B. Gadway, Exploring quantum signatures of chaos on a Floquet synthetic lattice, Phys. Rev. A {\bf 100}, 013623 (2019).

\bibitem{ZYLi2022}
Z. Y. Li, S. Colombo, C. Shu, G. Velez, S. Pilatowsky-Cameo, R Schmied, S. Choi, M. Lukin, E. Pedrozo-Ped{\~n}afiel, and V. Vuleti{\'c}, Improving Metrology with Quantum Scrambling, Science {\bf 380}, 1381 (2023).

\bibitem{Garttner18}
M. G\"{a}rttner, P. Hauke, and A. M. Rey, Relating Out-Of-Time-
Order Correlations to Entanglement Via Multiple-Quantum Coherences, Phys. Rev. Lett. {\bf 120}, 040402 (2018).

\bibitem{Zhao23pra201}
W. L. Zhao, R. R. Wang, H. Ke, and J. Liu, Scaling laws of the out-of-time-order correlators at the transition to the spontaneous PT-symmetry breaking in a Floquet system, Phys. Rev. A {\bf 107}, 062201 (2023).

\bibitem{HZhao21}
H. Zhao, Inferring the dynamics of ``black-box'' systems using a learning machine, Science China, Physics, Mechanics and Astronomy {\bf 64}, 270511 (2021).

\bibitem{Mano21}
T. Mano and T. Ohtsuki, Machine learning the dynamics of quantum kicked rotor, Ann. Phys. {\bf 435}, 168500(2021).

\bibitem{Lemarie10JMO}

G. Lemari{\' e}, D. Delande, J. C. Garreau and P. Szriftgiser, Classical diffusive dynamics for the quasiperiodic kicked rotor J. Mod. Opt. {\bf 57} 1922šC7, 2010.


\bibitem{Wlzhao23sym}
W. L. Zhao and H. Q. Zhang, Dynamical stability in a non-Hermitian kicked rotor model, Symmetry {\bf 15}, 113 (2023).

\bibitem{Wywang22}
W. Y. Wang and W. L. Zhao, Protected quantum coherence by gain and loss in a noisy quantum kicked rotor, J. Phys. Condens. Matter {\bf 34}, 025403 (2022).

\bibitem{KQHuang21pra}
K. Q. Huang, W. L. Zhao, and Z. Li, Effective protection of quantum coherence by a non-Hermitian driving potential, Phys. Rev. A {\bf 104}, 052405 (2021).

\bibitem{KQHuang21jpcm}
K. Q. Huang, J. Z. Wang, W. L. Zhao and J. Liu, Chaotic dynamics of a non-Hermitian kicked particle, J. Phys.: Condens. Matter {\bf 33}, 055402 (2021).

\end{thebibliography}
\end{document}